\documentclass{PoS}

\PoS{PoS(jhw2004)013}

\title{Matter and Gravity in Warped Extradimensional Models: Reinterpreting Randall-Sundrum}

\ShortTitle{Matter and Gravity in Warped Extradimensional Models: Reinterpreting RS}

\author{\speaker{Katherine Benson}\\
Emory University Physics Department\\
Atlanta, Georgia 30322-2430, USA\\
        E-mail: \email{benson@physics.emory.edu}}

\abstract{Warped extra dimension claims remarkable success: solving
 the hierarchy problem; explaining hierarchies in particle
 phenomenology; yielding standard cosmology, plus interesting
 nonstandard scenarios. Yet it has marked shortcomings: we over-rely
 on a single toy model, Randall-Sundrum; we treat matter and gravity
 in an ad hoc, asymmetric way; and we conceptualize integrated 4D
 effective field theory inconsistently. I here construct sounder 4D
 effective field theories for matter and gravity in warped extra
 dimension --- whether Randall-Sundrum or higher codimension. I track
 {\em both} Planck and particle scales through brane formation,
 beginning with fully extradimensional matter and gravity, at unified
 scale, in gravitationally warped backgrounds with bulk electroweak
 symmetry-breaking. This validates hierarchy solution, as warp
 generically drives 4D effective Planck and particle scales apart. It
 evades classic obstacles to warped confinement of matter:
 colocalizing particles and assuring effective charge
 universality. Diverse particles do fail to colocalize, generically
 and in discussed models; however, aggregate 4D effective field theory
 still holds, since hierarchy solution fixes unresolvably small
 extradimensional radius. Electromagnetic charge universality emerges
 generically, and weak charge universality in the Randall-Sundrum
 case.}

\FullConference{28th Johns Hopkins Workshop on Current Problems in Particle Theory\\
		 June 5-8, 2004\\
		 Johns Hopkins University Homewood campus - Bloomberg Center for Physics and Astronomy, Baltimore, Maryland}

\begin{document}

\section{Introduction: The origins of warped extra dimension}

In the past century, theorists' attempts to reconcile gravity and
particle interactions have involved a recurrent theme, of extra
spatial dimensions in our universe.  This notion arose in the
1920s, when Kaluza and Klein tried to model electromagnetism as
gravitational dynamics of a compactified extra dimension.\cite{klein,
kaluza} It reappeared in supergravity, string and M theory: unified
treatments of quantum gravity and particle interactions, consistent
{\em only} in higher dimension.  Early theorists viewed the required
extra dimensions as unresolvably small, with no measurable impact on
observed 4D physics.  Brane solutions to string theory --- defects
confining matter, but not gravity, through stringy effects
\cite{polchinski, duff} --- changed that view.  With matter confined
to 4D branes, dilution of gravity in {\em large} extra dimensions
could explain its hierarchical weakness, without violating 4D
observation. \cite{largeextra} This dilution scales with volume, for
flat extra dimensions. However, as defects, branes typically carry
tension, gravitationally warping the extra dimensions.  In 5D toy
models, Randall and Sundrum showed this gravitational warp can
accomplish two things: binding massless gravitons, to induce 4D
effective gravity on a brane; \cite{rs2} and generating hierarchy
between apparent Planck and particle scales, on a (different) brane.
\cite{rs1} Thus warped extra dimension emerged in 1999, as a new
reconciler of gravity and particle interactions, rendering both
effectively 4D and inducing their hierarchy from a single
extradimensional scale.

The Randall-Sundrum models established a paradigm for warped extra
dimension, wherein the apparent dimension of our universe
dynamically shifts to 4, when branes form, simultaneously binding
matter and gravity, and generating their hierarchy. This
paradigm sparked extensive research: both Randall-Sundrum papers
amassed more than 2000 citations, and ranked in the top 10 for high
energy physics citations each year since 2000, according to the SLAC
SPIRES database. This has two reasons. First, the Randall-Sundrum
models are toy models par excellence: simple, calculable, with deep
connections to fundamental paradigms like braneworld physics, AdS-CFT
issues, and the Horava-Witten vacuum in string theory. Second, their
application generated phenomenal success, in both phenomenology and
cosmology. Within cosmology, for example, they reproduce standard
Friedmann-Robertson-Walker universe evolution for late times, in
one-brane \cite{Binetruy,Csaki,Cline,Flanagan} and two-brane variants
\cite{Lukas, Cline2, Davis}; alter inflationary physics in acceptable
and potentially interesting ways \cite{Maartens, Copeland, Himemoto};
or seed late-time quintessence.  \cite{Langlois, Davis} These
successes can be transformative for particle cosmology.

I note that such success is not unparalleled: flat models of extra
dimension generate a comparable literature of phenomenological and
cosmological achievements, particularly in two models, the original
large extra dimensions proposal \cite{largeextra} and a subsequent
``infinite volume'' extra dimensions proposal. \cite{DGP} However,
warped extra dimension has more natural appeal.  For large,
``infinite volume,'' and warped extradimensional models are all
intrinsically braneworld models, deriving their viability from the
dynamical confinement of ordinary matter on branes. Yet branes
typically carry nonzero tension, which itself {\em generically} warps
spacetime.

\section{Critique: Shortcomings in the paradigm}

The success of warped extra dimension, in solving the hierarchy
problem; explaining hierarchies within particle physics; replicating
4D gravity and standard big bang cosmology; and seeding inflationary
and quintessential dynamics is spectacular.  However, serious
shortcomings persist, motivating the work I discuss today. These
shortcomings have two origins: first, over-reliance on the
Randall-Sundrum model to represent the warped extradimensional
paradigm; and second, ad hoc and asymmetric treatment of matter and
gravity effective theories in warped extradimensional backgrounds.

The first shortcoming stems simply from overreach.  For all their
promise, phenomenological and cosmological results nearly all arise in
one narrow context: Randall-Sundrum's toy models, with simple
extensions (bulk scalar fields --- themselves necessary to stabilize
Randall-Sundrum I \cite{Goldberger}; bulk gauge fields \cite{Pomarol,
Carena}; adding cosmological energy density to Randall-Sundrum
backgrounds \cite{Langlois}, and to detuned Randall-Sundrum
backgrounds, with $\Lambda_4 \ne 0$ \cite{Langlois}).  Warped extra
dimension arises generically from braneworld physics; the
Randall-Sundrum models, however, need not. They model a well-motivated
vacuum solution to M theory, the Horava-Witten vacuum, with branes at
fixed points of a large orbifolded extra dimension. \cite{hwitten} Yet
the Horava-Witten vacuum is one string vacuum, among a multitude of
contenders. \cite{KKLT,Susskind,Douglas,Ashok,Banks,Denef} Moreover,
the very role of string vacua may be limited by finite temperature
effects.  Brandenberger and Vafa address dimensional reduction of
string theory from this finite temperature perspective, asking in
thermodynamic equilibrium how dimensions {\em grow}.  For toroidal
compactifications they find a dimension-selecting mechanism, with
persistent cycle windings obstructing expansion in all but a
dimensionally reduced subspace.  \cite{bbergervafa} Recent analysis
suggests that often 4 dimensions grow very large, while 2 grow to
intermediate scale. \cite{bberger2, easther} Such arguments motivate,
among myriad possibilities, scenarios with 2 large extra dimensions,
warping when 3-branes form.  Thus a codimension 2 warped
extradimensional model, like \cite{Cohen, Chodos, Vilenkin, Gregory,
Shaposhnikov1, Ponton, Giovannini, Gregory2, deCarlos1, deCarlos2,
btoebbe}, might act as final stage, completing the chain of dynamical
dimensional reduction from 6 to 4, just as Randall-Sundrum completes
dimensional reduction to 5 to 4 after formation of a Horava-Witten
vacuum.  It is the general paradigm of warped extra dimension --- more
broadly than the Randall-Sundrum model --- which dimensionally reduces
our observed universe, when 3-branes form, warp spacetime, and induce
effectively 4D observable physics.  This warped braneworld mechanism
completes a dynamical pattern of dimensional reduction; one whose
pathways, however, are many.  Warped extra dimension thus deserves
{\em generic} study, examining its implications for 4D effective
physics, for novel phenomenology and cosmology, not through the
limited lens of a single toy model, but with an eye toward capturing
the generic, defining the landscape of structures and signatures
inexorably associated with warped extra dimension, regardless of
model.

Some progress toward this more generic perspective has occurred.  A
promising direction, that of extradimensional defect models, returns
to the inspiration of branes as defect solutions required by duality.
These construct branes explicitly as field theoretic, topological
defects, formed by an uncharged matter condensate in the extra
dimensions. The defect warps the metric, which then binds gravitons
to a 3-brane at the defect's core. These various solutions
retain the basic advantages of the Randall-Sundrum model: they confine
gravity to a brane; produce a massless bound graviton mediating
effective 4D gravity; cause only mild gravitational corrections via
Kaluza-Klein modes; and solve the hierarchy
problem. \cite{Cohen,Chodos,Gregory,Shaposhnikov1, Ponton, Giovannini,
Gregory2,DeWolfe,Shaposhnikov2,Randjbar-Daemi,Moon,bensoncho,btoebbe}
They also diversify the warped extradimensional universe paradigm,
suggesting a plethora of models with varying codimension for the
warped extra dimensions. Friedmann equations and brane cosmology have
been examined in codimension two, reproducing --- for a single model,
with thickened branes --- standard Friedmann-Robertson-Walker
limits. \cite{Cline4} These generalizations enable us to begin
separating generic from model-dependent features of warped extra
dimension.

However, these generalizations suffer from the second shortcoming of
warped extradimensional physics, shared by the Randall-Sundrum model
itself. This is an ad hoc and asymmetric treatment of gravity and
matter 4D effective field theory. These warped extradimensional
models all {\em derive} localization of massless gravitons on the
brane, and the resulting shift in Planck scale, from a fundamental
extradimensional, to an induced 4D effective, value. Most however {\em
assume}, a priori, localization of ordinary matter on the brane.  All
--- even the most ambitious --- {\em assume} an effective scale for
the 4D localized matter, equating it to the extradimensional Planck
scale. Historically this arose for two reasons: first, string
theoretic arguments compel confinement of ordinary particles to the
brane, where strings end; while confining gravity requires a distinct,
non-stringy mechanism. Offering this mechanism was itself the first
success of warped extra dimension. Second, a more symmetric treatment,
confining matter to branes via the same gravitational warp that
confines gravitons, is an enterprise with a long history.  Despite
some success \cite{Rubakov1,Rubakov2,Shifman}, longstanding problems
remain: first, in confining gauge fields; and second, in confining
matter while preserving 4D effective charge universality.
\cite{Rubakov} Thus most authors apply the mechanism of
gravitationally warped confinement {\em only} to gravitons.  Those who
attempt a parallel gravitational confinement of matter confine it only
as defect zero modes; that is, as {\em massless}
particles. \cite{Oda,Randjbar2} This is an impressive achievement, in
establishing field theory mechanisms to implement stringy matter
confinement on branes. However, it abstains on a question central to
hierarchy solution: establishing particle mass scales on the brane.

As I argue in \cite{warprs}, this asymmetric treatment of 4D effective
Planck and particle scales is untenable.  However it is
conceptualized, it is the single event of brane formation which
simultaneously reduces {\em both} matter and gravity to effective
4D theories. Our warped extradimensional paradigm derives a sometimes
marked shift in gravity's effective scale during that localization;
yet it assumes the particle scale stays fixed.  Solving the hierarchy
problem requires more: a closer, more consistent accounting, tracking
{\em both} particle and Planck scales from the same starting point,
through brane formation, when both scales evolve to their effective 4D
values in the nontrivial context of brane-warped geometry.  Such
tracking is needed, not only to validate hierarchy solution, but also
to validate brane cosmology and phenomenology.  For all rely centrally
on calibrated particle interactions on the brane; interactions whose
parameters may shift nontrivially in brane-warped backgrounds.

This asymmetric approach is exacerbated in the Randall-Sundrum
two-brane model, which employs an eclectic treatment of integrated 4D
effective field theory for gravity.  4D effective gravity has support
localized on a single brane; however Randall-Sundrum treat its 4D
effective action as salient, and equal, everywhere in the bulk.
Moreover, they assume this equal bulk action remains locally resolved:
with precisely this local resolution --- this recasting of an
integrated bulk action in terms of local warp factors --- responsible
for the generation of {\em distinct} 4D Planck scales on different
branes. This attribution of an integrated action fully and
simultaneously to all distinguishable points, without regard to the
action's support at each, is murky; more dubious is the pointwise
resolution of an already integrated effective action. A more
conventional treatment would regard the integrated 4D effective action
for gravity as either 1) localized on the 4D brane of its support; or
2) a bulk effect, describing an aggregate consequence of an extra
dimension itself too tiny to be resolved.

\section{Program: Constructing valid braneworld effective theories}

The work I discuss today addresses these shortcomings of warped extra
dimension, as practiced. First, I broaden the universe of viable
warped extradimensional models beyond the Randall-Sundrum example,
establishing {\bf generic} features and structures. Second, I ratchet
upward the standard for viability. I deem warped extradimensional
models viable if 1) they determine a single, well-defined, 4D
effective field theory for matter and gravity, which fully and
symmetrically encodes all consequences of extradimensional warp; and
2) they induce hierarchy of particle and Planck scales, within that
valid 4D effective theory.

Fundamental to a consistent and symmetric treatment of matter and
gravity is a computational strategy to peel back the veil: to
extend the history of {\em both} particle and Planck
scale from today's 4D values backward, through their dynamical
confinement onto 4D branes, to an earlier unified extradimensional
scale. Viewed in forward rather than reverse, this requires consistent
tracking of particle and Planck scales forward: from a single unified
starting point, through brane formation, when both scales evolve to
effective 4D values in the brane-warped background.

The only calculable tool for such tracking is extradimensional field
theory: assuming particles have a definite, field theoretic
description not only after they are confined to the 3-brane, but
before, in an extradimensional universe. Such extradimensional
effective field theories need not arise in string theory; however,
they can, and when they do, they provide a definite technical
framework for calculating the shift in effective parameters ---
including particle and Planck scales --- when branes form and mediate
spontaneous dimensional reduction of the universe. This technical
framework is completely prescriptive: it offers neither need nor
permission for ad hoc assumption, but enables instead a complete and
consistent derivation of consequences of extradimensional warp, for
{\em both} matter and gravity. Gravity, we've seen, can have its
effective Planck scale shifted markedly from the fundamental string
scale, due to extradimensional warp. Extradimensional field theory for
matter lets us probe how the same extradimensional warp affects the
effective particle scale: whether it too necessarily shifts, or
remains --- as previously assumed ad hoc --- at the fundamental
string scale.

To explore this possible transmutation of the particle scale, I begin
with a fully extradimensional field theory with negative bulk
cosmological constant, in which {\em both} gravity and matter are
inherently bulk fields with unified mass scale. The fundamental ---
extradimensional -- particle scale arises from bulk electroweak
symmetry-breaking, and coincides with the fundamental ---
extradimensional --- Planck scale. Brane formation establishes a
warped background metric, which in turn induces integrated 4D Kaluza
Klein effective field theories for {\em both} matter and gravity. Note
that I take {\em all} matter fields to be bulk fields; {\em all}
physics flows from extradimensional to 4D, controlled by the
brane-warped background metric. I calculate the induced 4D effective
field theories for both gravity and matter; establish their
sensibility; and derive the effective 4D particle and Planck scales
they imply, to rigorously validate or refute hierarchy solution.

\section{Results: Valid 4D effective field theories with hierarchy}

In recent work I apply this extradimensional field theoretic
approach to construct viable warped braneworlds, with hierarchy, in
three contexts: first, for generic warped extradimensional models of
arbitrary codimension \cite{warpxd}; second, for specific example
extradimensional defect models of codimension two \cite{btoebbe} and
three \cite{bendomokos}; and third, for the warped geometry of the
Randall-Sundrum form \cite{warprs}. In each, I find the central
motivation of warped extra dimension, hierarchy generation, rigorously
validated; however, the well-defined 4D effective field theories
obtained are interpretable only as either 4D aggregate theories, which
encode effects of unresolvably small bulk dimension; or potentially,
as localized 4D theories on a single brane. The formalism supports
only these interpretations --- not the Randall-Sundrum lore of a
universal bulk effective theory valid, and differentially
resolvable, at each point in the bulk.

In \cite{warpxd} I examine generic warped extradimensional models, of
diverse warp and codimension, with bulk electroweak symmetry-breaking
from a constant Higgs condensate throughout the bulk.  I calculate
integrated 4D effective theories for both matter and gravity in the
brane-warped background explicitly: deriving for gravity, and for
scalars, fermions and gauge fields, their 4D Kaluza-Klein reductions.
I find generic results: the effective 4D graviton is {\em always} a
constant zero mode solution, whose normalization {\em always} fixes
the induced 4D Planck mass. The effective 4D photon is {\em always} a
constant gauge zero mode, which {\em always} couples universally to
electromagnetically charged fields. 4D electromagnetic charge
universality thus follows {\em automatically} from warped extra
dimension, in any model with appropriate bulk electroweak
symmetry-breaking.  Localization profiles {\em always} differ for
massless gravitons, photons, massive gauge bosons and massive
fermions. This clarifies the structural challenge in constructing
warped spacetimes where gravity fully colocalizes entire 4D effective
theories on a single brane. However, integrated 4D effective theory
still holds for {\em small} warped extra dimension, as an
unresolved, aggregate theory. Hierarchy solution remains
model-dependent, but structurally favored; indeed, the structure of
warped extra dimension offers a blueprint for hierarchy
generation. For Planck and particle scales are set by distinct
physical mechanisms: the Planck scale, by normalization of the
graviton zero mode; the particle scale, by lowest Kaluza-Klein masses
for Higgs and weak gauge fields. Divergence of these distinct flows
occurs readily, inducing hierarchy. Moreover, imposing hierarchy
solution typically fixes the allowed extradimensional radius,
determining whether a particular model supports the small warped extra
dimension scenario.

More concretely, for the most general spherically symmetric warped metric in $D$ dimensions, with $\Lambda_D <0$,
\begin{equation}
ds^2      = B(r)\  \ \bar{\!\!\!\!g\,\,\,}_{\mu\nu}\ dx^\mu dx^\nu  +A(r)\ dr^2 + r^2d\Omega^2\ \ ,
\end{equation}
we may have a flat braneworld solution $\ \bar{\!\!\!\!g\,\,\, }_{\mu\nu} = \eta_{\mu\nu},\ \Lambda_4 = 0$, with background warp factors $A(r)$, $B(r)$. In extradimensional defect models, these warp factors $A(r)$, $B(r)$ are fixed by an uncharged extradimensional defect condensate, with defect core at $r=0$.

Graviton fluctuations $\ \bar{\!\!\!\!g\,\,\, }_{\mu\nu} = \eta_{\mu\nu} + \bar{h}_{\mu\nu}$ have modulating bulk wave functions $\bar{h}_n(r)$ which obey (in the $l=0$ case) 
\begin{equation}
\frac{1}{BA^{1/2}r^{D-5}}\ \ \frac{d}{dr}\
\left( \ B^2A^{-1/2}r^{D-5} \ \frac{d\bar{h}_n}{dr} \right)  \ =\ 
- \, m_n^2\, \bar{h}_n\, .
\end{equation}
This is a Sturm-Liouville equation for the graviton mode $\bar{h}_n$, with weight $\rho
= B\, A^{1/2}r^{D-5}$ and 4D effective Kaluza-Klein mass
$m_n$.  The lowest eigenvalue $m_n=0$, a graviton zero mode whose regular solution is simply the constant $\bar{h}_o = 1/\sqrt{N}$, mediating a 4D effective $1/r^2$ gravitational interaction. This mode has spatial probability distribution
\begin{equation}
\rho |\bar{h}_o|^2 \sim B\,
A^{1/2}r^{D-5}\, .
\end{equation}  
Thus the warped metric solution for $A(r)$ and $B(r)$ directly  dictates
the graviton zero mode's spatial profile: whether, where, and how
sharply it localizes in the bulk. The normalization factor $N$ determines the 4D effective Planck mass:
\begin{equation}
M_{\rm Pl, 4}^2 = N\ M_{\rm Pl, D}^{D-2} \ \ .
\end{equation}

Scalars and fermions with bulk electroweak mass $m_{D}$
have modulating bulk Kaluza Klein wave functions $H_n(r)$ which obey (again in the $l=0$ case)
\begin{equation}
 \frac{1}{BA^{1/2}r^{D-5}}\ \ \frac{d}{dr}\
\left( \ B^2A^{-1/2}r^{D-5} \ \frac{dH_n}{dr} \right)
\ -\  B\, m_{D}^2\  H_n \ =\ -  \, m_n^2\, H_n\, .
\end{equation}
These have 4D Kaluza Klein masses $m_n$ and Sturm-Liouville weight, as in the gravitational case,  $\rho
= B\, A^{1/2}r^{D-5}$.  The lowest Kaluza Klein mass $m_1 > 0$ fixes the 4D effective particle scale, with spatial  probability distribution
\begin{equation}
\rho |{H}_1|^2 \sim B\,
A^{1/2}r^{D-5}\, |{H}_1|^2\ \ .
\end{equation}  
This differs from the massless graviton's spatial profile by the nontrivial eigenfunction $|{H}_1|^2$, and so can be localized differently within the bulk.

Gauge bosons with bulk electroweak mass $m_{a, D}$
have modulating bulk Kaluza Klein wave functions ${\cal A}_{an}(r)$
which obey (again in the $l=0$ case)
\begin{equation}
 \frac{1}{A^{1/2}r^{D-5}}\ \ \frac{d}{dr}\
\left( \ BA^{-1/2}r^{D-5} \ \frac{d{\cal A}_{an}}{dr} \right)
\ -\ B\, m_{a,D}^2\  {\cal A}_{an} \ =\ - \, m_{an}^2\, {\cal A}_{an}\,  ,
\end{equation}
given the transverse gauge choice
\begin{equation}
0 \ =\  \sum_{{\mbox{{\tiny transverse\ }}\atop X}}  \nabla_X\, (B^{-1}A_a^X) \ \ \ ,
\end{equation}
which exhausts gauge freedom in the extra (transverse) dimensions
only. These have 4D Kaluza Klein masses $m_{an}$ and Sturm-Liouville weight
$\rho_{\cal A}\ =\ A^{1/2}r^{D-5} \ = \ B^{-1}\,\rho_H$
where $\rho_H$ gives the weight for gravitons, scalars, and fermions.

For the photon, with $m_{a, D} = 0$, there is again a single regular
massless mode, the constant ${\cal A}_{ao} = 1/\sqrt{N_{\cal
A}}$. This gauge zero mode becomes the effective 4D photon, with
spatial profile
\begin{equation}
\rho_{\cal A} |{\cal A}_{ao}|^2 \sim \, 
A^{1/2}r^{D-5}\, 
\end{equation} 
distinct from the massless graviton's profile $BA^{1/2}r^{D-5}$.
Photons and $1/r^2$ gravity thus {\em generically} have distinct
localization profiles in the bulk, profiles modulated  by the
nontrivial warp factor  $B(r)$. While this might
conceivably colocalize photons and 4D gravity on a single
brane, in a specific warped model, the generic expectation must be
noncoincident distribution --- making a single 4D integrated effective
theory for matter and gravity valid {\em only} as an aggregate effective
theory, for unresolvably small  warped extra dimension.  

For the massive weak bosons, the lowest Kaluza Klein mass $m_{a1} > 0$ fixes the 4D effective gauge masses, with spatial profile 
\begin{equation}
\rho_{\cal A} |{\cal A}_{a1}|^2 \sim \, 
A^{1/2}r^{D-5}\, |{\cal A}_{a1}|^2
\end{equation} 
This differs not only from graviton, scalar, and fermion spatial profiles, but also from the photon's spatial profile, by the nontrivial eigenfunction $|{\cal A}_{a1}|^2$. Thus we have generic and rampant diversity in which the 4D effective particles from  four classes --- massless bulk gravitons, scalars and fermions; massive bulk scalars and fermions; bulk photons; and massive bulk weak gauge bosons --- each localize differently within the bulk.

We see here the structural predisposition to hierarchy generation ---
due to the distinct physical mechanisms driving the 4D effective
Planck mass to $M_{\rm Pl, 4}^2 = N\ M_{\rm Pl, D}^{D-2}$, and the 4D
effective particle masses to $m_1$, $m_{a1}$ for the Higgs and weak gauge
bosons. Detailed examination of this divergence, however, remains
model-dependent, as  specific warped solutions $A(r),\ B(r)$ fix
both $N$ and the Kaluza Klein masses $m_1$, $m_{a1}$:  $N$ quite directly, but $m_1$ and $m_{a1}$ in roundabout, complex ways. 

With Domokos \cite{bendomokos} and with Toebbe \cite{btoebbe}, I
establish numerical solutions for two specific warped extradimensional
defect models, with 4D induced matter effective theories, localized
gravity, and hierarchy solution all rigorously validated. These
involve, in the first case, 4D matter effective theory in the
Benson-Cho extradimensional global monopole model \cite{bensoncho}; in
the second, a new warped solution and 4D effective theory for matter
and gravity, for an extradimensional global string.  In both cases,
hierarchy solution remains robust, as Planck and particle scales
diverge hierarchically. For both, the Planck scale grows
hierarchically, while the particle scale shifts little: in the
monopole case, flowing to or remaining near a natural value of order
$|\Lambda_7|$. Thus both are characterized by unified extradimensional
matter and gravity at the {\rm TeV} scale, inducing 4D Planck scale
gravity and 4D {\rm TeV} scale matter. Both involve diversely
localized fields, with 4D integrated effective theories valid only for
extradimensional radii too small to resolve; yet both numerically
accommodate this possibility.

In \cite{warprs}, I apply this same extradimensional field theoretic
approach to the Randall-Sundrum warped background. I view both matter
and gravity as inherently 5D, as bulk fields with unified mass scale
--- with 5D particle scale again set by bulk electroweak
symmetry-breaking. Here, however, technicalities arise, as
$Z_2$-orbifold symmetry requires bulk fermion masses to be $Z_2$-odd
in the bulk, assuming a kink profile $m {\rm sign}(y)$. This is
accomplished by a $Z_2$-odd Higgs vev, with comparable kink profile,
yielding uniform masses for gauge and Higgs particles. These yield
integrated Kaluza-Klein effective theories for matter which have
appeared in the literature before (for gravity, in \cite{rs1}; for
bulk scalars, in \cite{Goldberger2}; for bulk gauge fields, in
\cite{Pomarol}; and for bulk fermions, in \cite{gpomarol, ngrossman,
chang}). In my paradigm, though, these effective matter fields are
again the {\em only} matter fields; {\em all} fields are bulk fields,
and {\em all} physics flows from the bulk to 4D.

This unified extradimensional approach rigorously validates hierarchy
solution, in a more constrained interpretation than Randall-Sundrum:
unified 5D matter and gravity at the Planck scale induces 4D Planck
scale gravity and 4D {\rm TeV} scale matter, when an extra dimension
of size $12\ M_{\rm Pl}^{-1}$ warps due to brane formation. Our
formalism sheds the ambiguity latent in Randall-Sundrum: the unified
bulk scale {\em must} lie at the Planck mass, as the effective 4D
Planck mass changes little, while the effective 4D particle mass-scale
becomes hierarchically suppressed. Note that this instantiates a
distinct avenue to warp-generated hierarchy, unlike the
extradimensional defect models above, where unified bulk scale at a
{\rm TeV} induces a little-changed 4D particle scale and a
hierarchically enhanced 4D Planck scale.  In each case, a hierarchical
divergence arises between induced 4D particle and Planck scales; yet
whether that divergence arises from enhancement of the Planck scale
(as in the extradimensional defect models) or a suppression of the
particle scale (as for the Randall-Sundrum background) remains
model-dependent.

For the Randall-Sundrum background, this extradimensional field
theoretic approach automatically solves a {\em second} problem: the
dimensionally reduced matter effective field theory preserves both
electromagnetic and weak charge universality.  This offers a rare
achievement: a field theoretic dimensional reduction with demonstrable
weak charge universality.  Moreover, while fields localize
differentially in the bulk --- massless gravitons on the Planck brane;
massive matter fields on the particle brane; photons smeared
throughout the bulk --- the integrated 4D effective field theory
remains sensible for all fields. This is because
hierarchy demands such a small warped extra dimension --- $12 M_{\rm
Pl}^{-1}$ --- that varied localization in the extra dimension remains
unresolved. The observationally valid 4D effective
theory is the smeared, aggregate one obtained by integration.  Thus we
have in the Randall-Sundrum case a viable warped model, one which: 1)
induces a single dimensionally reduced 4D theory of clear validity for
both matter and gravity; and 2) generates hierarchy, while preserving
both electromagnetic and weak charge universality. Yet this viable theory describes not the
large and resolvable bulk interpreted by Randall-Sundrum, but an aggregate 4D
theory due to a truly tiny --- but warped --- extra dimension.

\section{Conclusions}

I introduced the warped extradimensional paradigm, then critiqued its
shortcomings: overdominance of a single toy model (Randall-Sundrum);
asymmetric treatment of matter and gravity 4D effective field
theories; and eclectic spatial interpretation of integrated effective
field theories.

I then defined a unified extradimensional theory for matter and
gravity, in a generic warped extradimensional background with bulk
electroweak symmetry-breaking. I examined Kaluza-Klein reduction in
this warped background, deriving integrated 4D effective field
theories for gravity, and for free scalars, fermions, and gauge
fields. I discussed induced 4D Planck and particle scales, and
localization of diverse modes.  I noted that massless modes have
universal form, determining distinct localization profiles for
gravitons and photons. This generically obstructs  construction of
warped spacetimes with fully colocalized local 4D effective
theories. Hierarchy solution allows an alternative: smeared 4D
effective theories, from unresolvably small extra dimension. Because
the divergence between induced particle and Planck scales typically
scales with extradimensional radius, hierarchy solution fixes, for
each model, a scenario of definite large or small extra dimension. In
either case --- whether colocalized or smeared over unresolvably small
extra dimension --- whenever the induced 4D effective field theories
are sensible, they generically generate hierarchy and preserve electromagnetic
charge universality.

I described specific warped extradimensional defect models, which
induce valid 4D effective field theories with unresolvably small extra
dimension and rigorously validated hierarchy solution. In them a
unified bulk scale at a {\rm TeV} induces 4D {\rm TeV} scale matter
and Planck scale gravity. In the Randall-Sundrum background, instead,
unified 5D matter and gravity originates at the Planck scale, inducing
4D Planck scale gravity and 4D {\rm TeV} scale matter when an extra
dimension of unresolvably tiny size $12\ M_{\rm Pl}^{-1}$ warps.  This case, as well, yields  well-defined 4D effective field theories, with rigorously validated hierarchy solution,
electromagnetic and weak charge universality.

\end{document}